\documentclass[final, 5p, twocolumn, 10pt, a4paper, times, sort&compress]{elsarticle}

\pdfoutput=1

\makeatletter
\def\ps@pprintTitle{%
  \let\@oddhead\@empty
  \let\@evenhead\@empty
  \def\@oddfoot{\reset@font\hfil\thepage\hfil}
  \let\@evenfoot\@oddfoot
}
\makeatother

\usepackage[cmex10]{amsmath}
\usepackage{algorithmic}
\usepackage{array}
\usepackage{url}
\usepackage{natbib}
\usepackage{siunitx}
\usepackage{amsmath, amssymb, bm}
\usepackage[english]{babel}
\usepackage[final, colorlinks]{hyperref}
\usepackage[percent]{overpic}

\hypersetup{pdftitle={Gradient-based quantitative image reconstruction in ultrasound-modulated optical tomography: first harmonic measurement type in a linearised diffusion formulation}}

\hypersetup{pdfauthor={Samuel Powell, Simon R. Arridge, and Terence S. Leung}}

\hypersetup{pdfsubject={Quantitative ultrasound-modulated optical tomography}}

\hypersetup{pdfkeywords={Ultrasound-modulated optical tomography, acousto-optic effects, inverse problems, finite element analysis, gradient methods}}

\let\oldhat\hat
\renewcommand{\vec}[1]{\mathbf{#1}}
\renewcommand{\hat}[1]{\oldhat{\mathbf{#1}}}

\newcommand{\plusplus}{\nolinebreak\hspace{-.05em}\raisebox{.4ex}{\tiny\bf +}\nolinebreak\hspace{-.10em}\raisebox{.4ex}{\tiny\bf +} }

\newcommand{\kth}{k^{\textrm{th}}}

\newcommand{\sakadzic}{Sakad\v{z}̌i\'{c} }

\newcommand{\qa}{{q_1(\vr)}}

\newcommand{\phiz}{\phi(\vr)}
\newcommand{\vphiz}{\bm{\phi}}
\newcommand{\phidc}{\phi_0(\vr)}

\newcommand{\phid}{\phi_\delta(\vr, \tau)}
\newcommand{\phib}{\phi_b(\vr, \tau)}
\newcommand{\phia}{\phi_1(\vr)}
\newcommand{\vphia}{\bm{\phi}_1}
\newcommand{\phip}{\bm{\phi}^+}

\newcommand{\uoteff}{\eta(\vr)}
\newcommand{\veta}{\bm{\eta}}
\newcommand{\diffopp}{\mathcal{L}(\kappa, \mua)}
\newcommand{\vn}{\vec{n}}
\newcommand{\vr}{\vec{r}}

\newcommand{\vy}{\vec{y}_1}
\newcommand{\vx}{\vec{x}}
\newcommand{\vym}{\vec{y}_1^m}
\newcommand{\drm}{\,\textrm{d}}
\newcommand{\vb}{\vec{b}}
\newcommand{\meas}{\mathcal{D}}
\newcommand{\vmeas}{\vec{D}}
\newcommand{\cmua}{\mathcal{C}_1^{\mua}}
\newcommand{\cmusp}{\mathcal{C}_1^{\musp}}

\newcommand{\xvar}{\bm{x}}
\newcommand{\xgvar}{\bm{x}_g}

\newcommand{\fxvar}{\tilde{{x}}}

\newcommand{\ige}{\Gamma_e^{-1}}
\newcommand{\mua}{\mu_a}
\newcommand{\musp}{\mu_s'}
\newcommand{\proj}{\mathcal{P}}
\newcommand{\vproj}{\vec{P}}
\newcommand{\forward}{\mathcal{F}}

\newcommand{\real}{\mathbb{R}}
\newcommand{\normal}{\mathcal{N}}
\newcommand{\efunc}{\mathcal{E}(\xvar)}
\newcommand{\defunc}{\mathcal{E}'(\xvar)}
\newcommand{\vefunc}{\vec{E}(\fxvar)}
\newcommand{\dvefunc}{\vec{E}'(\fxvar)}

\newcommand{\dr}{\,\textrm{d} \vr}
\newcommand{\basis}[1]{u_{#1}(\vr)}

\newcommand{\sysmat}{\vec{S}}

\newcommand{\midx}[1]{\{#1\}}
\newcommand{\vbs}{\vb^*}

\newcommand{\ka}{k_{\mathrm{a}}}
\newcommand{\ltr}{l_{\mathrm{tr}}}

\begin{document}
\begin{frontmatter}

\title{Gradient-based quantitative image reconstruction in ultrasound-modulated optical tomography: first harmonic measurement type in a linearised diffusion formulation}

\author[cs]{\corref{cor1}Samuel~Powell}
\ead{s.powell@ucl.ac.uk}
\author[cs]{Simon~R.~Arridge}
\author[medphys]{Terence S. Leung}
\cortext[cor1]{Corresponding author}

\address[cs]{Department of Computer Science, University College London, Gower Street, LONDON, WC1E 6BT, UK.}
\address[medphys]{Department of Medical Physics and Biomedical Engineering, University College London, Gower Street, LONDON, WC1E 6BT, UK.}

\begin{abstract}
Ultrasound-modulated optical tomography is an emerging biomedical imaging modality which uses the spatially localised acoustically-driven modulation of coherent light as a probe of the structure and optical properties of biological tissues. In this work we begin by providing an overview of forward modelling methods, before deriving a linearised diffusion-style model which calculates the first-harmonic modulated flux measured on the boundary of a given domain. We derive and examine the correlation measurement density functions of the model which describe the sensitivity of the modality to perturbations in the optical parameters of interest. Finally, we employ said functions in the development of an adjoint-assisted gradient based image reconstruction method, which ameliorates the computational burden and memory requirements of a traditional Newton-based optimisation approach. We validate our work by performing reconstructions of optical absorption and scattering in two- and three-dimensions using simulated measurements with 1\% proportional Gaussian noise, and demonstrate the successful recovery of the parameters to within $\pm5$\% of their true values when the resolution of the ultrasound raster probing the domain is sufficient to delineate perturbing inclusions.
\end{abstract}


\end{frontmatter}

\section{Introduction}
\label{sec:introduction}
The wavelength-dependent optical absorption and scattering coefficients of biological tissues provide clinically valuable information regarding tissue function and composition. Purely optical techniques such as diffuse optical tomography (DOT) are capable of measuring these coefficients but suffer from limited spatial resolution due to the high degree of optical scattering encountered in typical biological media \cite{Arridge:2009cy,Arridge:1999kd,Gibson:2005hr}. Ultrasound-modulated optical tomography (UOT) is a hybrid technique which aims to recover the coefficients with significantly improved resolution by combining the optical contrast of near infra-red light with the spatial resolution of focused or time-gated ultrasound fields.

Much effort has been expended in advancing the experimental technique in UOT. The problem of detecting the small ultrasound-modulated optical flux against the large unmodulated background has received significant attention. This problem is particularly challenging since the requisite use of a coherent source generates a spatially incoherent speckle pattern on the boundary of the domain. The flux of individual coherence areas must therefore be collected in parallel \cite{Gross:2003ei, Leveque:1999tm, Li:2002wq}, or manipulated such that their contributions can be measured in summation \cite{Ramaz:2004vs,Murray:2004tb,Bossy:2005tr,Li:2008hy,Suzuki:2013ft}. 

In addition to the spatial incoherence of the generated speckle pattern, the Brownian motion present in living tissues causes temporal decorrelation which can further complicate the experimental technique. Some methods demonstrate inherent immunity, such as the direct digital autocorrelation of the detected speckle field \cite{Li:2002jg}, and spectral hole burning \cite{Li:2008hy}. Other methods which employ holographic techniques, such as the use of photo-refractive crystals \& polymers \cite{Ramaz:2004vs,Murray:2004tb,Suzuki:2013ft}, require that the response time of the medium is faster than the decay rate of the tissues (in the order of milliseconds).

Less attention has been paid to the fundamental problem that hybrid techniques such as UOT are only capable of producing quantitative images under some form of model-based reconstruction procedure. This was succinctly demonstrated by the images produced by Lev and Sfez \cite{Lev:2002ba, Lev:2003dt} where the raw data from a UOT measurement can be seen to resemble the optical sensitivity functions \cite{Arridge:1995fy} in a given domain. 

Previous investigations have successfully recovered the optical absorption coefficient from simulated \cite{Allmaras:kr,Powell:2013dx} and experimental data  \cite{Bratchenia:2011ch}. Simultaneous recovery of both the optical absorption and scattering coefficients raises the more subtle problem of non-uniqueness. The key idea is that the recovery of the two coefficients requires at least two sets of \emph{internal} data, as has been demonstrated in a related incoherent formulation of UOT \cite{Bal:2010gv, Bal:2014cg}. We recently demonstrated that uniqueness can be restored by the use of multiple optical source and detector locations \cite{Powell:2014eu}.

UOT is often compared with photo-acoustic tomography (PAT), another hybrid method which exploits the limited scattering of ultrasound in biological tissues to enhance the spatial resolution of recovered \emph{optical} properties. In PAT a pulsed laser illuminates the tissue, and regions of optical absorption undergo thermo-elastic expansion, generating an ultrasound wave which is detected on the surface of the medium \cite{Beard:2011bm}. Techniques such as acoustic time-reversal are then used to reconstruct the original absorbed energy distribution. Measurements in PAT are thus principally sensitive to the absorption coefficient inside the medium, whereas UOT is sensitive to both absorption and scattering perturbations \cite{Kothapalli:2007hx}. PAT can achieve higher transverse spatial resolution than UOT, as it is not dependent on the ability to focus an acoustic field to a particular point in the medium, though the achievable resolution is typically depth dependent as high frequency components in the measured acoustic field are attenuated by tissue \cite{Wang:2008ci}. Both techniques require reconstruction methods to quantitatively map the optical properties of tissues \cite{BenCox:2012jr,Gao:2012cu,Saratoon:2013bm,Hochuli:2015et}.  Accordingly, the practical capabilities of PAT and UOT can be seen as largely complimentary: UOT offers the potential to achieve the recovery of absorption and scattering with millimetric resolution, at significant depth, and PAT more readily achieves sub-millimetre resolution in the absorption coefficient at smaller depths.

\subsection{Overview and contribution}
This work is organised as follows. In section \ref{sec:theory} we provide an overview of forward modelling techniques for UOT, and derive an efficient model of the power-spectral density of the UOT signal from a non-linear time-domain form presented elsewhere in the literature. In section \ref{sec:inverseproblem} we pose the inverse problem of recovering the internal optical parameter distributions from measured data, deriving the correlation measurement density functions which describe the sensitivity of our measurements to pertubations in the parameters of interest. In section \ref{sec:implementation} we demonstrate the implementation of our techniques by the finite-element method. We employ our reconstruction methods in section \ref{sec:results} by performing, for the first time, simultaneous reconstruction of the optical properties in a simulated UOT experiment in two- and three-dimensions. We close this work with a discussion of our findings in section \ref{sec:conclusion}.

\section{Forward model}
\label{sec:theory}
\subsection{The physical basis of UOT}

As an acoustic wave propagates through a biological medium it induces small changes in the refractive index of the medium \cite{Wang:2001jd}, and causes the displacement of optical scatterers from their rest position \cite{Leutz:1995hf, Kempe:1997ig}. Under coherent illumination an otherwise static (in the absence of Brownian motion) speckle pattern is generated on the surface of the medium which changes in time as the optical path lengths of the scattered waves travelling through the medium are phase modulated by the acoustic field. These changes may be measured as either a temporal decorrelation of the intensity autocorrelation function, or, equivalently, as the modification of the power spectral density of the measured light \cite{Elson:tm}.

\subsection{Forward modelling techniques}

Various models of this process have been presented in the literature. The starting point in each case is a time-domain description of the phase perturbations applied to the scattered optical waves propagating through the medium. 

\subsubsection{Path-integral methods} develop expressions for the total average phase perturbation over an optical path of a given length by averaging the phase perturbation expression over all free-paths and scattering events. Under an assumption of weak scattering the perturbations applied to each optical path length are considered independent, and each provides an individual contribution to the total field autocorrelation function. Integration of the contributions to the correlation function over a probability distribution of path lengths, typically found analytically from the diffusion equation, yields an estimate of the measured field autocorrelation function. This approach is similar to some of the original investigations into diffusing wave spectroscopy by Maret and Wolf \cite{Maret:1987vl}, and Pine et al. \cite{Pine:1988vp}.

The principal limitation of path integral methods is that they can only incorporate planar acoustic fields (owing to the averaging over all potential paths). Despite this limitation, the technique has offered significant insight into UOT for time-harmonic acoustic fields in isotropically \cite{Wang:2001jd} and anisotropically scattering \cite{Sakadzic:2002bj} media, and for acoustic pulses \cite{Sakadzic:2005go} in which there are significant correlations in the phase perturbations between successive scattering events. 

\subsubsection{Correlation transport} is an extension of radiative transport theory (a high-frequency approximation for optics) to consider media in which there is a temporal variation of the underlying medium. This idea was originally investigated by Ackerson et al. \cite{Ackerson:1992fx} who proposed a \emph{correlation} transport equation (CTE) for use in the field of diffuse correlation spectroscopy (DCS).  Dougherty et al. \cite{Dougherty:1994vn} later provided a more rigorous derivation based upon analytic theory with moving scatterers, building on the work of Ishimaru and Hong \cite{Ishimaru:1975uv} and others \cite{Stephen:1988in,MacKintosh:1989jz}. A similar approach can be taken in UOT \cite{Sakadzic:2007is,Sakadzic:2006ei} where the scattering objects now move deterministically, and additional terms are introduced to account for the modulation of the refractive index. 

The only necessary assumption in a CTE based model is that of weak scattering ($\ltr \gg \lambda$, where $\ltr = (\mua + \musp)^{-1}$ is the optical mean free path, $\mua$ is the optical absorption coefficient, $\musp = \mu_s(1-g)$ is the reduced scattering coefficient, $g$ is the scattering anisotropy, and $\lambda$ is the optical wavelength). Owing to the complexity and high-dimensionality of the phase space of the integro-differential form of the CTE in UOT, solutions have to date only been presented using statistical approximations sought via the computationally expensive Monte-Carlo (MC) method. We have previously presented GPU implementations of a CTE for UOT \cite{Leung:2010jr, Powell:2012jb} which demonstrate significant improvements in speed over traditional implementations, but even so, the computation requirements are such that at present these models are only suitable for use as a `gold-standard' by which more approximate techniques may be validated.

\subsubsection{Correlation diffusion} is an approximation of correlation transport by a first order expansion of a given CTE in spherical harmonics, valid under a number of assumptions, principally that $\mu_s' \gg \mu_a$. The diffusion approximation is readily made for the RTE in the context of DOT, and also the CTE which arises in DCS, see Boas \cite{Boas:1997kf} for a full derivation. The process is more complicated in UOT, principally due to correlations between phase increments at successive scattering sites. By combining aspects of the statistical averaging of the path-integral methods with the transport component of the CTE, \sakadzic and Wang \cite{Sakadzic:2006tx} developed a correlation diffusion equation (CDE) for the field autocorrelation function in UOT.

Approximations in the derivation of the CDE presented in [40] require two conditions over those assumed in the derivation of the CTE. The first is that the phase increments between successive scattering events are uncorrelated ($\ka \ltr \gg 1$, where $\ka$ is the acoustic wavenumber). The second is that of moderate ultrasound pressures (circa $10^5$Pa for typical medical ultrasound frequencies). The flexibility and computational simplicity of this model are highly attractive for application in an image reconstruction method. 

\subsection{Linearised power-spectral correlation diffusion model}

The CDE presented in ref. \cite{Sakadzic:2006tx} describes the wide-sense stationary optical field autocorrelation function $\phi(\vr, \tau)$ in a given domain, and it can be written:
\begin{align}
\left [\diffopp + \eta(\vr) \left (1-\cos(\omega_a \tau) \right ) \right ] \phi(\vr, \tau) &=  \label{eq:cdenonlinbody}\\
\phi(\vr, \tau) + 2 A \kappa \vn \cdot \nabla \phi(\vr, \tau) &= q^-(\vr)
\label{eq:cdenonlinbody2}
\end{align}
with a lossy diffusion operator
\begin{equation}
\diffopp = -\nabla \cdot \kappa \nabla + \mua,
\end{equation}
where $\kappa = (3 \musp)^{-1}$, is the diffusion coefficient, $q^-(\vr)$ is a coherent boundary flux source, $A$ is related to the index of refraction mismatch between the turbid and external media \cite{Schweiger:1995to},  $\vec{n}$ is a unit vector normal to the boundary of the domain, $\omega_{\mathrm{a}}$ is the acoustic angular frequency, $\tau$ is lag, $\eta(\vr) \propto P_0(\vr)^2$ is the acousto-optic modulation efficiency, which has weak dependence upon the optical parameters, and $P_0(\vr)$ is the local acoustic pressure amplitude.

In previous work \cite{Powell:2013dx} we employed this approximation as part of a linear reconstruction technique for UOT, neglecting the weak dependence upon optical coefficients in $\eta(\vr)$, successfully recovering absorption coefficients from boundary measurements. This model is readily applicable to techniques in which the intensity or field autocorrelation function are directly recorded in the temporal lag domain. However, many interferometric detection methods in UOT directly record the power-spectral density of outgoing modulated flux at the acoustic frequency (the first harmonic flux). To calculate this value multiple runs of the forward model must be made at difference values of lag, $\tau$. Given that the magnitude of $\eta(\vr)$ is already limited by the assumption of weak modulation, we are motivated to further linearise the model of eq.\ (\ref{eq:cdenonlinbody}) to arrive at a direct frequency (power-spectral) domain representation.

Suppose we take two measurements of the same medium under two acoustic pressures such that the field correlation function $\phi(\vr, \tau) \mapsto \phib + \phid$, under $\eta \mapsto \eta_b + \eta_\delta$, then
\begin{multline}
\left [\diffopp + (\eta_b + \eta_\delta) \left(1 - \cos(\omega_a \tau)\right) \right]\\
\times [\phi_b(\vr, \tau) + \phi_\delta(\vr, \tau)] = 0.
\label{eq:cdenonlinpert1}
\end{multline}
Subtracting from eq.\ (\ref{eq:cdenonlinpert1}) the expression for the `baseline' measurement made under $\eta_b$ results in a non-linear expression for the perturbed field under changing insonification,
\begin{multline}
\diffopp  \phid
+ \eta_b \left(1 - \cos(\omega_a \tau)\right)  \phid
\\+  \eta_\delta(1-\cos(\omega_a\tau))\phid
=- \eta_\delta(1-\cos(\omega_a\tau))\phib.
\label{eq:nonlinearderiv}
\end{multline}

In eq.\ (\ref{eq:nonlinearderiv}) we identify the second order terms in the small parameter $\eta_\delta$, which we neglect to complete our linearisation:
\begin{multline}
\diffopp \phid  
+ \eta_b \left(1 - \cos(\omega_a \tau)\right)  \phid 
\\=- \eta_\delta(1-\cos(\omega_a\tau)) \phib.
\label{eq:eqformlinearise}
\end{multline}
This form has significant value since judicious choices of ultrasound pressures and a known pressure squared dependence may permit $\eta(\vr)$ to be identified experimentally. For now, we choose our background measurement to be made in the absence of an acoustic field such that $\eta_b = 0$, and $\eta_\delta = \eta$. Inserting these definitions and rewriting the expression as a set of coupled equations, 

\begin{align}
\diffopp \phib &= 0, \label{eq:cdelinbase}\\
\diffopp \phid &=- \eta(1-\cos(\omega_a\tau))\phib\label{eq:cdelinpert}.
\end{align}

By inspection, it is evident that the linearised field autocorrelation function $\bar{\phi}(\vr, \tau) = \phib + \phid$ contains spectral content only at DC and the frequency of the ultrasonic excitation $\omega_{\textrm{a}}$. We may now find an expression for the power-spectral density of the fluence in the domain by the Wiener-Khinchin theorem. Taking the Fourier transform of the linearised field autocorrelation function,
\begin{equation}
\mathcal{F} [\bar{\phi}(\vr,\tau)] (\omega) = \hat{\phi}(\vr, \omega).
\end{equation}
We refer to the fluence at the ultrasonic frequency as the first harmonic correlation fluence $\phia = \hat{\phi}(\vr, \omega_{\textrm{a}})$, and the total fluence rate, equivalent to the CW fluence in a DOT experiment,
\begin{equation}
\phiz = \int_{-\infty}^{\infty} \hat{\phi}(\vr, \omega) \drm \omega = \phi(\vr, 0).
\end{equation}
Performing the Fourier transform we find,
\begin{align}
\diffopp \phia &= \uoteff\phi(\vr) = \qa,\label{eq:linsystemphia}\\
\diffopp \phiz &= 0, \label{eq:linsystemphiz}\\
\phia + 2 A \kappa \vn \cdot \nabla \phia &= 0\label{eq:linsystembnda},\\
\phiz + 2 A \kappa \vn \cdot \nabla \phiz &= q^-(\vr)\label{eq:linsystembndz},
\end{align}
The components at the fundamental, $\phidc = \phiz-\phia = \hat{\phi}(\vr, 0)$, and first harmonic, $\phia$, both contain spatially localised information related to the total fluence rate in the medium by the presence of the modulation efficiency term. In this model we may readily identify the oft-cited `virtual acousto-optic source' as the product $\uoteff\phiz$. 

We define our measurement as the power spectral \emph{flux} across the boundary at the ultrasound frequency. For a point detector this is given by \cite{Arridge:1999kd},
\begin{equation}
y_1(\vr) = -\kappa \vn \cdot \nabla \phia,\hspace{.5cm}  \vr \in \delta \Omega.
\end{equation}

Measurements of the first harmonic flux are a natural data type produced by interferometric, holographic, or spectral-hole burning instrumentation in the specified regime. This data type can also be readily calculated from measurements of the autocorrelation of the flux exiting the domain by application of a Fourier transform.

Defining an extended measurement aperture $d(\vr), r \in \delta \Omega$, and combining the expression of the flux with the Robin boundary condition allows us to define a measurement operator,
\begin{equation}
\mathcal{D}[d(\vr)]\phi_1(\vr) = \int_{\delta\Omega} \frac{1}{2 A} d(\vr) \phia \dr = \left< m, \phia \right>_{\delta \Omega}.
\end{equation}

Before proceeding to consider the inverse problem, we make two comments on our model. First, we note that if $\eta(\vr) = \delta(\vr - \vr_0)$, the `internal data' returned by a UOT measurement is in fact a measurement of the diffuse optical absorption sensitivity function for a pointwise perturbation in the domain at $\delta(\vr_0)$ (though this is not in practice possible due to the inherently finite nature of a practically realisable ultrasound field distribution). We will return to this point in the discussion. Second, a model similar to that of equations \ref{eq:cdelinbase} and \ref{eq:cdelinpert} was previously developed by Allmaras and Bangerth \cite{Allmaras:kr} by the application of the Born approximation to the modulated field in a path integral formulation, the nature of those paths then being formalised in the diffusion framework (similar to earlier works by \cite{Sakadzic:2002bj}). Our derivation shows that this model can be derived by reasoned approximations to a correlation transport equation, and that it is then amenable to a power-spectral representation.

\subsection{Measurement protocol and notation}

To this point we have considered the case of a single optical source and detector, and ultrasound field distribution. To perform imaging we will probe the domain of interest with multiple insonification profiles and optical source-detector pairs. We define a measurement index 
 $\rho(i,j,k) = k + N_j\times j + N_j\times N_i \times i$ for optical sources $i = 1,\ldots,N_i$, detectors $j=1,\ldots,N_j$, and acoustic field distributions $k=1,\ldots,N_k$.

For a given experiment the data vector $\vy \in \real^{N_m}, N_m = N_i \times N_j \times N_k$ contains data from all combinations of sources, detectors and acoustic field profiles. The data vector is found by application of a  stacked set of projection operators which correspond to all combinations of source, detector and acoustic field, $\bm{\proj} = \begin{pmatrix} \proj_1, \ldots, \proj_{N_m} \end{pmatrix}^T$. The stacked projection operator applies the identically ordered measurement operator $\bm{\mathcal{D}}$ to a forward operator, $\bm{\forward}$, which implements eq.\ (\ref{eq:linsystemphia}) under parametrisation by the optical parameters of interest.

To refer to a single measurement the subscript $\rho$ implies a particular choice of the 3-tuple $\{i,j,k\}$ which corresponds to the appropriate subset of the forward operators, and potentially their derivatives. That is to say that,
\begin{align}
y_{1, \rho = \{i,j,k\}} = \proj_{\{i,j,k\}}[\mua,\musp] = \meas_{\{j\}} \forward_{\{i,k\}} = \meas[d_{\{j\}}] \forward_{\{i,k\}},\\
\forward_{\{i,k\}} = \diffopp \phi_{1, \{i,k\}} = \eta_{\{k\}} \phi_{\{i\}},\\
\diffopp \phi_{\{i\}} = q^-_{\{i\}}.
\end{align}
where we have now dropped the spatial dependence of the various fields and apertures. For convenience we also denote the compound optical parameters ${\xvar} = \left( \mua, \musp \right)^T$.

\section{The inverse problem}
\label{sec:inverseproblem}
To determine the parameters of our reconstructed image $\xvar_*$ we take a regularised output least squares approach, corresponding to the minimisation of the error functional \cite{Arridge:2009cy}
\begin{equation}
\bm{x}_* =\underset{\bm{x}}{\operatorname{argmin}} \,\efunc := \frac{1}{2} \|\vym - \bm{\proj}[\xvar] \|^2_{\Gamma_e^{-1}} + \frac{\lambda}{2} \mathcal{R}[\xvar -\xgvar],
\label{eq:efunc}
\end{equation}
where we assume our measurements $\vym = \vy + \vn$ are corrupted by noise $\vn \sim \normal(0, \Gamma_e)$ drawn from a normal distribution with zero mean and covariance $\Gamma_e$, $\mathcal{R}$ is a suitable regularisation operator, the hyper-parameter $\lambda$ serves to control the relative contribution of the data term and the regularisation term to the error functional, and $\bm{x}_g$ is an a priori reference parameter set. The nature of $\bm{x}_g$ could, for example, be determined from approximate measurements of the domain through alternative modalities. In our work we assume no such additional knowledge, such that $\bm{x}_g = 0$ throughout.

\subsection{The error functional gradient}

Under the assumption that the error functional is convex, minimisation of $\efunc$ corresponds to the solution of the non-linear equation $\defunc = 0$. To find this expression we begin by expanding equation \ref{eq:efunc},
\begin{equation}
\efunc = \frac{1}{2} \left(\vym - \bm{\proj}[\xvar]\right)^T\Gamma_e^{-1}\left(\vym - \bm{\proj}[\xvar]\right) 
+ \frac{\lambda}{2}  \mathcal{R}[\xvar_\Delta],
\label{eq:efuncexp}
\end{equation}
where $\xvar_\Delta = \xvar -\xgvar$. Denoting the residual $\vb = \left(\vym - \bm{\proj}[\xvar]\right)$ and taking the derivative with respect to $\xvar$,
\begin{equation}
\defunc = -\bm{\proj}'^*[\xvar] \Gamma_e^{-1}  \vb  +  \lambda  \mathcal{R}'[\xvar_\Delta],
\label{eq:derefunc}
\end{equation}
where $\bm{\proj'}$ and $\mathcal{R}'$ are the Fr\'echet derivatives of the projection operator and regularisation operators respectively, and the superscript $^*$ denotes the adjoint.

\subsection{Sensitivity functions}
The Fr\'echet derivative of the projection operator is a linear mapping which describes changes in our measurement resulting from perturbations in the optical parameters. To define the operator we consider, for a particular optical source, detector, and acoustic field, the perturbations $\phi \mapsto \phi + \phi^{\delta}$, $\phi_1 \mapsto \phi_1 + \phi_1^{\delta}$ which occur under $\mu_{\rm a} \mapsto \mu_{\rm a} +\mu^{\delta}_{\rm a} $, $\kappa \mapsto \kappa + \kappa^{\delta}$. After dropping second order terms we have from eq.\ (\ref{eq:linsystemphia}) that
\begin{equation}
\mathcal{L}(\kappa,\mu_{\rm a}) \phi^{\delta}_1 =\eta  \phi^{\delta} -\mathcal{L}(\kappa^{\delta},\mu^{\delta}_{\rm a})\phi_{1}\label{eq:linsystemphia_pert},
\end{equation} 
and from eq.\ (\ref{eq:linsystemphiz}) that
\begin{equation}
\mathcal{L}(\kappa,\mu_{\rm a}) \phi^{\delta} = -\mathcal{L}(\kappa^{\delta},\mu^{\delta}_{\rm a})\phi.\label{eq:linsystemphiz_pert}
\end{equation}
Application of the measurement operator to the perturbed first harmonic field gives the value of the Fr\'echet derivative of the first harmonic correlation flux evaluated at some $\kappa$, $\mu_{\rm a}$,
\begin{equation}
y_{1. \rho}^{\delta} =  \left< m, \phi^{\delta}_1 \right>_{\partial \Omega}.
\label{eq:y1pert}
\end{equation}
Our task is to find an expression for this inner product which can be calculated efficiently. Defining the DC adjoint field
\begin{equation}
\mathcal{L}^{*}(\kappa,\mu_{\rm a})\phi^{+} = m,
\label{eq:adjfield0}
\end{equation} 
we proceed from eq.\ (\ref{eq:y1pert}):
\begin{align}
 \left< m, \phi_1^{\delta}\right>_{\partial \Omega} &=  \left<\mathcal{L}^{*}(\kappa,\mu_{\rm a})\phi^{+} , \phi_1^{\delta}\right>_{\partial \Omega}\nonumber \\
 &=  \left<\phi^{+} ,\mathcal{L}(\kappa,\mu_{\rm a}) \phi_1^{\delta}\right>_{\Omega}\nonumber \\
 &=  \left<\phi^{+} ,\eta  \phi^{\delta} -\mathcal{L}(\kappa^{\delta},\mu^{\delta}_{\rm a})\phi_1\right>_{\Omega}\nonumber \\
 &=  \left<\eta \phi^{+} , \phi^{\delta}\right>_{\Omega} - \left<\phi^{+} ,\mathcal{L}(\kappa^{\delta},\mu^{\delta}_{\rm a})\phi_1\right>_{\Omega}\label{eq:splitadj_both}\,.
\end{align}
The second term on the right hand side of eq.\ (\ref{eq:splitadj_both}) involves the inner product of the first harmonic correlation fluence and the adjoint DC fluence from eq.\ (\ref{eq:adjfield0}). Inserting the lossy diffusion operator, assuming $\kappa^{\delta} = 0$ on the boundary, and applying the divergence theorem,
\begin{align}
\left<\phi^{+} ,\mathcal{L}(\kappa^{\delta},\mu^{\delta}_{\rm a})\phi_1\right>_{\Omega} 
 &= \int_{\Omega} \mu^{\delta}_{\rm a} \phi^{+}  \phi_1 - \phi^{+} \nabla \cdot \kappa^{\delta}  \phi_1 \\ 
 &= \int_{\Omega} \mu^{\delta}_{\rm a} \phi^{+}  \phi_1 + \kappa^{\delta}  \nabla \phi^{+} \cdot \nabla \phi_1. 
\label{eq:splitadj2}
\end{align}
To develop the first term on the right hand side of eq.\ (\ref{eq:splitadj_both}) we define the first harmonic adjoint field,
\begin{equation}
\mathcal{L}^{*}(\kappa,\mu_{\rm a})\phi_1^{+} = \eta \phi^{+}\label{eq:adjfield1}\,,
\end{equation} 
whence
\begin{align}
 \left<\eta \phi^{+} , \phi^{\delta}\right>_{\Omega} &=  \left<\mathcal{L}^{*}(\kappa,\mu_{\rm a})\phi_1^{+}  , \phi^{\delta}\right>_{\Omega}\nonumber \\
 &= \left<\phi_1^{+} , \mathcal{L}(\kappa,\mu_{\rm a}) \phi^{\delta}\right>_{\Omega}\nonumber \\
 &= - \left<\phi_1^{+} , \mathcal{L}(\kappa^{\delta},\mu^{\delta}_{\rm a}) \phi\right>_{\Omega}\,.
\end{align}
This term involves the inner product of the DC fluence and the adjoint first harmonic fluence from eq.\ (\ref{eq:adjfield1}). As with eq.\ (\ref{eq:splitadj2}) we can employ the divergence theorem to state this term explicitly under the same assumptions:
\begin{align}
\left<\phi_1^{+} ,\mathcal{L}(\kappa^{\delta},\mu^{\delta}_{\rm a})\phi\right>_{\Omega} 
&= \int_{\Omega} \mu^{\delta}_{\rm a} \phi_1^{+}  \phi - \phi_1^{+} \nabla \cdot \kappa^{\delta}, \\ 
&= \int_{\Omega} \mu^{\delta}_{\rm a} \phi_1^{+}  \phi +  \kappa^{\delta}  \nabla \phi_1^{+} \cdot \nabla \phi. 
\label{eq:splitadj1}
\end{align}
The value of the mapping defined by the Fr\'echet derivative of the projection operator is thus given by,
\begin{equation}
y_{1,\rho}^{\delta} = - \int_{\Omega}  \mu^{\delta}_{\rm a} \left ( \phi_1^{+}  \phi  +  \phi^{+}  \phi_1 \right)  +  \kappa^{\delta}  \left ( \nabla \phi_1^{+} \cdot \nabla \phi + \nabla \phi^{+} \cdot \nabla \phi_1 \right ).
\label{eq:ydeltafinal}
\end{equation}
Noting the linearity of eq.\ (\ref{eq:ydeltafinal}) in the optical parameters, we define the total derivatives of our measurement with respect to the parameters of the forward model \cite{Arridge:1999kd},
\begin{align}
\frac{\mathrm{d} y_{1, \rho}}{\mathrm{d} \mua}  = \cmua    &=  - \left( \phi_1^{+}  \phi  +  \phi^{+}  \phi_1\right), \label{eq:ccontmua}\\
\frac{\mathrm{d} y_{1, \rho}}{\mathrm{d}  \musp} = \cmusp &= 3 \kappa^2 \left( \nabla \phi_1^{+} \cdot \nabla \phi + \nabla \phi^{+} \cdot \nabla \phi_1 \right ) \label{eq:ccontmusp}.
\end{align}
Consistently with our previous work \cite{Powell:2013dx}, we refer to these functions as the first harmonic \emph{correlation measurement density function} (CMDF), in analogy with the associated \emph{photon measurement density functions} arising in DOT \cite{Arridge:1995fy,Arridge:1995ho}. We will consider the form of these functions further in section \ref{sec:results}. The pairs of CMDFs for the complete set of measurements form the kernel of the Fr\'echet derivative of the projection operator, a continuous to discrete linear mapping from perturbations in the optical parameters to changes in the measurement,
\begin{equation}
\vec{y}_{1}^{\delta}  = \bm{\proj}'[x] x^{\delta}  =  \int_\Omega \begin{pmatrix} \bm{\mathcal{C}}^{\mua}_{1} & \bm{\mathcal{C}}^{\musp}_{1} \end{pmatrix} \begin{bmatrix} \mua^{\delta} \\ \musp^{\delta} \end{bmatrix} \dr.
\end{equation}
where $\bm{\mathcal{C}}^{\mu}_{1} = (\mathcal{C}^{\mu}_{1,1}, \ldots, \mathcal{C}^{\mu}_{1,N_m})^T$ for $\mu = \mua$, $\musp$, is a set of stacked CMDFs over each combination of source, detector and acoustic field. The adjoint Fr\'echet derivative of the forward operator, required in the error functional gradient, is a discrete to continuous linear operator which back-projects changes in measurements on the boundary to the parameter space,
\begin{equation}
\vec{h}(\vr) = \begin{pmatrix} h^{\mua} & h^{\musp} \end{pmatrix}^T = \bm{\proj}'^*[x] \vec{y}_1^{\delta} =\begin{pmatrix} \bm{\mathcal{C}}^{\mua}_{1} & \bm{\mathcal{C}}^{\musp}_{1} \end{pmatrix}^T \vec{y}_{1}^{\delta} .
\label{eq:adjderproj}
\end{equation}

\section{Implementation}
\label{sec:implementation}

To proceed we must now redefine our problem in a finite-dimensional setting suitable for solution using numerical methods, and decide upon a technique by which we solve the large non-linear system resulting from the discretisation of the expression $\defunc=0$.

\subsection{Finite element implementation of the forward model}

The domain under consideration is subdivided into a mesh of non-overlapping elements joined at $N_n$ vertex nodes. On this mesh we define a set of piecewise linear basis functions such that $u_i(\vr_j) = \delta_{ij}$ for $i,j = 1, \ldots, N_n$ where $\vr_j$ located at the $j^{\textrm{th}}$ vertex node. We may then define finite dimensional approximations to the parameter distributions, solutions, and sources,
\begin{equation}
\chi(\vr) \approx \tilde{\chi}(\vr) = \sum^{N_n}_k \chi_k \basis{k},
\end{equation}
where we take $\chi_k$ to represent the $\kth$ component of the $N_n \times 1$ vectors $\bm{\chi}$ of nodal coefficients which define the discrete approximation $\tilde{\chi}(\vr)$ of any of the continuous functions $\mua, \musp, \kappa, \eta, \phi_0, \phi_1, q_0, q_1$, and $\vec{x} = (\bm{\mua}, \bm{\musp})^T$ is the $2N_n \times 1$ compound vector of optical coefficients.

We solve the correlation diffusion equation by the finite element method \cite{Arridge:1993gv,Arridge:1995ho,Heino:2003bd}. Equation (\ref{eq:linsystemphia}) is multiplied by test functions which obey the boundary conditions, and whose zeroth and first derivatives are integrable over the domain. The boundary conditions of eqs.\ (\ref{eq:linsystembnda}) and (\ref{eq:linsystembndz}) are incorporated by subsequent integration by parts. The Galerkin formulation results from selecting the test functions in the weak formulation to be the same as the basis in which we have defined our parameters, and allows us to write the resulting linear system,
\begin{equation}
[\vec{K}+\vec{M}+\vec{F}]\vphia = \sysmat[\vx] \vphia = \vec{q}_1
\end{equation}
 where
\begin{align}
\vec{K} &= \sum_k \kappa_k \vec{V}_k^{\kappa}, && \vec{M} = \sum_k \mu_{a,k} \vec{V}_k^{\mua},
\end{align}
are the $N_n \times N_n$ matrices of system matrix integrals, formed by the basis system matrices
\begin{align}
{V}^{\kappa}_{k,ij} &= \frac{\drm \sysmat}{\drm \kappa}= \int_\Omega \basis{k} \,\nabla \basis{i} \cdot \nabla \basis{j}  \dr,
\label{eq:dsdkappa} \\
{V}^{\mua}_{k,ij} &= \frac{\drm \sysmat}{\drm \mua} =  \int_\Omega \basis{k} \basis{i} \basis{j}  \,\dr,
\label{eq:dsdmua}
\end{align}
and
\begin{align}
{F}_{ij} &=\frac{1}{2A}  \int_{\partial\Omega} \basis{i} \basis{j} \,\dr,\\
{q}_{1,i} &= \int_\Omega q_1(\vr) u_i(\vr) \, \dr\label{eq:deqdef}.
\end{align}

To find the vector of virtual source coefficients $\vec{q}_1$ we begin by identifying the discrete form of the right hand side of eq.\ (\ref{eq:linsystemphia}):
\begin{equation}
q_1(\vr) \approx \sum_i \eta_i \basis{i} \sum_j \phi_{0,j} \basis{j},
\end{equation}
which we insert into eq.\ (\ref{eq:deqdef}),
\begin{equation}
{q}_{1,k} = \sum_{i,j} \eta_i \phi_{0,j} \int_\Omega \basis{i} \basis{j} \basis{k} \, \dr = \veta^T \vec{V}^{\mua}_k \vphiz.
\end{equation}

The vector of coefficients $\vphiz$ employed in the definition of $\vec{q}_1$ themselves come from the solution of the system
\begin{equation}
\sysmat[\vx] \vphiz = \vec{q}^-
\end{equation}
which is the standard formulation of the continuous-wave DOT problem with source term coefficients found by projection of the boundary source profile into the basis. In this work we employed the Toast\plusplus toolbox to manage mesh data, peform the relevant elemental integrals, and assemble the system matrices and right-hand-sides \cite{Schweiger:2014ee}.

\subsection{Discrete error functional, and its gradient}

The discrete form of the error functional in eq.\ (\ref{eq:efuncexp}) is found by replacing the operators and fields with their discrete approximations,
\begin{equation}
\vefunc = \frac{1}{2}(\vym - \vproj[\fxvar])^T \ige (\vym - \vproj[\fxvar]) + \frac{\lambda}{2} \mathcal{R}[\fxvar_\Delta], 
\label{eq:defunc}
\end{equation}
where for a given measurement
\begin{equation}
\vec{P}_\rho[\fxvar] = \vec{D}^T_{\midx{j}} {\bm{\phi}_1}_{\midx{i,k}} = \vec{D}^T_{\midx{j}} \sysmat^{-1}[\fxvar] {\vec{q}_1}_{\midx{i,k}}.
\label{eq:dprojop}
\end{equation}
By the same procedure we find the discrete form of the derivative of the error functional of eq.\ (\ref{eq:derefunc}),
\begin{align}
\dvefunc & = -\vproj'[\fxvar]^T \ige \vb +  \lambda \mathcal{R}'[\fxvar_\Delta], \label{eq:dvefunc} \\
&= - \vec{J}[\fxvar]^T \ige \vb +  \lambda \mathcal{R}'[\vx_\Delta]. \label{eq:ddvefunc}
\end{align}
The derivative of the discrete projection operator $\vproj'[\fxvar] = \vec{J}[\fxvar]$ is the dense $N_m \times 2 N_n$ Jacobian matrix. Its transpose constitutes the discrete approximation to the adjoint derivative projection operator of eq.\ (\ref{eq:adjderproj}). Accordingly, $\vec{J}[\fxvar]$ consists of stacked pairs of $1 \times N_n$ correlation measurement density vectors $\vec{C}^{\mua}_{1,\rho}$, and $\vec{C}^{\musp}_{1, \rho}$, for each parameter $\mua$ and $\musp$, for every measurement $\rho$,
\begin{equation}
\vec{J}[\fxvar] = \begin{bmatrix}  \vec{C}_{1,1}^{\mua}  & \vec{C}_{1,1}^{\musp} \\
									\vdots & \vdots \\
									\vec{C}_{1,N_m}^{\mua}  & \vec{C}_{1,N_m}^{\musp} \end{bmatrix}.
\end{equation}
The discrete equivalents of the correlation measurement density functions of eqs.\ (\ref{eq:ccontmua}) and (\ref{eq:ccontmusp}) are:
\begin{align}
C_{1,\rho,i}^{\mua}  &= -\left(\vphiz^T {\vec{V}}^{\mua}_{i} \bm{\phi}_1^+ + \vphia^T {\vec{V}}^{\mua}_{i} \bm{\phi}^+ \right),\label{eq:cdescmua} \\
C_{1,\rho,i}^{\musp} &= 3{\tilde{\kappa}}_i^2 \left(\vphiz^T{\vec{V}}^{\kappa}_{i} \bm{\phi}_1^+ + \vphia^T {\vec{V}}^{\kappa}_{i} \bm{\phi}^+ \right) \label{eq:cdescmusp},
\end{align}
where we understand the fields to be those corresponding to measurement index $\rho$, the derivatives of the system matrix were defined in eqs.\ \ref{eq:dsdkappa} and \ref{eq:dsdmua}. Corresponding to the continuous forms of eqs.\ (\ref{eq:adjfield0}) and (\ref{eq:adjfield1}), the adjoint fields are found by solutions of the systems,
\begin{align}
\sysmat \phip = \vmeas, && \sysmat \bm{\phi}_1^+ = \bm{q}_{1}^+, && {q}_{1,k}^+ = \veta^T \vec{V}^{\mua}_k {\phip}. \label{eq:discadjs}
\end{align}

By employing the adjoint solutions to the forward problem we have derived a method by which the Jacobian can be calculated with at most four runs of the forward model for each measurement index, $\rho$. In a UOT experiment, it is highly unlikely that each $\rho$ will consist of an entirely unique set of optical sources, detectors, and acoustic field distributions. In this case pre-computation of all optical solutions and adjoints $\vphiz$ and $\bm{\phi}^+$ will result in significantly accelerated calculations. 

A typical approach to minimising $\vefunc$ is via a Newton based method such the Gauss-Newton and Levenberg-Marquardt techniques. Whilst such techniques avoid computation of the Hessian matrix by various approximations, the underlying Taylor expansion of the objective function still calls for the storage (and repeated inversion) of the complete Jacobian matrix. In techniques such as DOT there are typically a limited number of sources and detectors such that the viability of storing the Jacobian is principally dependent upon the number of degrees of freedom in the forward model, as determined by the finesse and dimension of the discretisation. This is also true of UOT, however in this application there may be an arbitrary number of measurements corresponding to different acoustic field profiles such that Jacobian matrix extends in both the row- and column-space. We therefore take an alternative approach which is to directly calculate the functional gradient on a row-by-row basis, and employ this in a gradient based optimisation technique which does not require storage of the explicit Jacobian. This technique has been previously demonstrated in DOT \cite{Arridge:1998wp} and quantitative photoacoustics \cite{Cox:2007he,Saratoon:2013bm}.

To proceed we incorporate the error covariance matrix into the residual such that $\vbs = \ige \vb$, and rewrite eq.\ (\ref{eq:ddvefunc}),
\begin{equation}
\dvefunc = -\sum_\rho \vproj_\rho'[\fxvar]^T  b^*_{\rho} +  \lambda \mathcal{R}'[\vx_\Delta].
\end{equation}
We define two adjoint fields modified from eq.\ \ref{eq:discadjs},
\begin{align}
\sysmat \bm{\phi}^*_{\rho} = \vmeas_{\rho}^T b^*_{\rho}, && \sysmat \bm{\phi}^*_{1,\rho} &= \bm{q}_1^*, && q_{1,k}^* = \veta^T \vec{V}^{\mua}_k \bm{\phi}^*_{\rho},
\end{align}
and the $N_n \times 1$ vectors
\begin{align}
z_i^{\mua} &= -\sum_\rho \left ( \bm{\phi}^T_{\rho} {\vec{V}}^{\mua}_{i} \bm{\phi}^*_{1,\rho} + \bm{\phi}^T_{1,\rho} {\vec{V}}^{\mua}_{i} \bm{\phi}^*_\rho  \right),\\
z_i^{\musp} &= 3 \tilde{\kappa}_i^2 \sum_\rho \left ( \bm{\phi}^T_{\rho} {\vec{V}}^{\kappa}_{i} \bm{\phi}^*_{1,\rho} + \bm{\phi}^T_{1,\rho} {\vec{V}}^{\kappa}_{i} \bm{\phi}^*_\rho \right),
\end{align}
such that,
\begin{align}
\dvefunc &= -(\vec{z}^{\mua}, \; \vec{z}^{\musp})^T + \lambda \mathcal{R}'[\vx_\Delta],\\
\dvefunc &= -\vec{z} + \lambda \mathcal{R}'[\vx_\Delta].
\label{eq:dvefuncz}
\end{align}
Equation (\ref{eq:dvefuncz}) permits the direct calculation of the gradient of the error functional without explicitly building the intermediary Jacobian matrix. 

\subsection{Regularisation}

In this work we employ first-order Tikhonov regularisation, which encourages smooth solutions to the inverse problem. Application of the regularisation operator and its derivative to the parameters in their discrete representation is therefore implemented as \cite{Schweiger:2005fa},
\begin{equation}
\mathcal{R}[\vx_\Delta] = \vec{x}_\Delta^T \vec{Y} \vec{x}_\Delta, \;\;\;
\mathcal{R}'[\vx_\Delta] = \vec{Y} \vec{x}_\Delta,
\end{equation}
where
\begin{equation}
\vec{Y} = \begin{bmatrix} {\bar{\mu}_{a}^{-2}} \vec{W} & \vec{0} \\ \vec{0} & {\bar{\mu}_{s}'^{-2}} \vec{W} \end{bmatrix}, \;\;\; {W_{i,j}} =  \int_\Omega \nabla \basis{i} \cdot \nabla \basis{j} \dr,
\end{equation}
and $\bar{\mu}_{a}$, $\bar{\mu}_{s}'$ are the mean values of the a priori reference parameters, $\mu_{\rm a,g}$, $\mu_{\rm s, g}'$, defined previously.

\subsection{Optimisation}
To solve the non-linear system $\dvefunc=0$ we employed a (Polak-Ribi\'{e}re) non-linear conjugate gradient method. The gradients were preconditioned with the block mass matrix:
\begin{align}
\begin{bmatrix} \vec{B} & \bm{0} \\ \bm{0} & \vec{B} \end{bmatrix}, \;\mathrm{ where }\; {B_{ij}} = \int_\Omega \basis{i} \basis{j} \dr.
\end{align}

For details of this optimisation technique, refer to standard texts, e.g., Nocedal and Wright \cite{Nocedal:428607}.

\section{Results}
\label{sec:results}
In this section we will examine the form of the CMDFs defining the sensitivity of our measurements, and demonstrate simulated reconstructions in two- and three-dimensions. 

Where we employ a two-dimensional domain, the three-dimensional formation of the theory is employed. Such `two and a half' dimensional settings imply continuity through the plane in the sources, optical properties, fields and sensitivity functions. That is to say that, for example, point sources implicitly represent infinitely extended line sources.

\subsection{Correlation measurement density functions}
\label{res:cmdfs}

The CMDFs which define the sensitivity of our measurement to perturbations in the optical properties of the system offer significant insight into this imaging modality. We will now consider the form of the individual CMDFs, under variation of the acoustic field and optical source-detector profiles.

\begin{figure}
\centering
\begin{overpic}{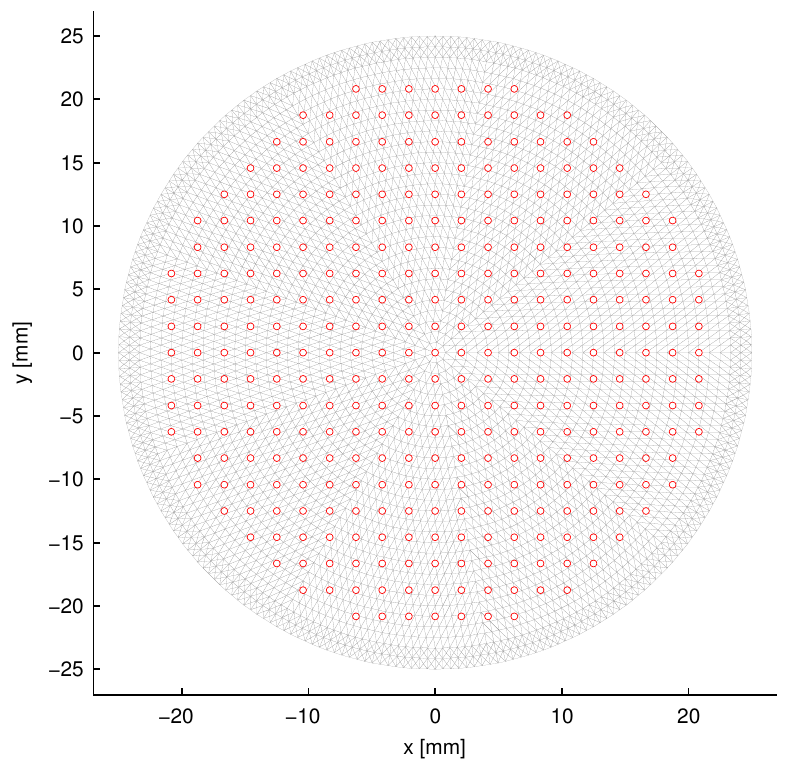}
\includegraphics{./fig/mesh2dfields}
\end{overpic}
\caption{Geometry of the two-dimensional mesh for CMDF inspection and reconstructions of section \ref{res:2drecon}. White dots indicate the location of the acoustic focii.}
\label{fig:mesh2dfields}
\end{figure}
For this purpose we consider a two-dimensional circular domain of diameter $50\si{\mm}$ with homogeneous optical properties. The absorption and reduced scattering coefficients $\mua = 0.01\si{\per\mm}$, $\musp = 1\si{\per\mm}$, of the domain are typical of biological tissues. The refractive index of the domain is matched to its surroundings. Around the periphery of the domain are placed three optical sources and three optical detectors two of which are collocated, each having Gaussian profiles of full-width half-maximum (FWHM) $5\si{\mm}$. Given the reciprocity of the problem there exist a set of six source-detector combinations which return unique data. The domain is discretised into a set of $6,840$ linear triangular elements joined at $3,511$ vertex nodes. A set of $349$ focused acoustic fields probes the domain through the two-dimensional plane. Each field has a Gaussian profile with FWHM = $2\si{\mm}$, and the set are arranged over a rectangular grid with spacing $4\si{\mm}$, truncated at a radius of $22\si{\mm}$ from the centre of the domain. The peak magnitude of $\eta(\vr) = 0.25$. The discretised domain is depicted in figure \ref{fig:mesh2dfields}, where each circle represents an acoustic focal point.

Expressions for the discrete form CMDFs in $\mua$ and $\musp$ were presented in eqs.\ (\ref{eq:cdescmua}) and (\ref{eq:cdescmusp}). Each expression consisted of four terms,
\begin{enumerate}
\item $\vphiz$, the forward total fluence, equivalent to the CW DOT fluence in the medium due to application of a given optical source.
\item $\vphia$, the first-harmonic UOT fluence which results from the virtual acousto-optic source given by the product of the light distribution from the optical source and the acousto-optic efficiency term.
\item $\bm{\phi}^+$, the adjoint total fluence, equivalent to the CW DOT fluence in the medium due to an adjoint source found by application of a given optical measurement operator.
\item $\bm{\phi}_1^+$, the adjoint first-harmonic UOT fluence which results from the adjoint virtual  acousto-optic source given by the product of the adjoint fluence distribution from the adjoint optical source and the acousto-optic efficiency term.
\end{enumerate}
\begin{figure*}[!htb]
\centering
\begin{overpic}{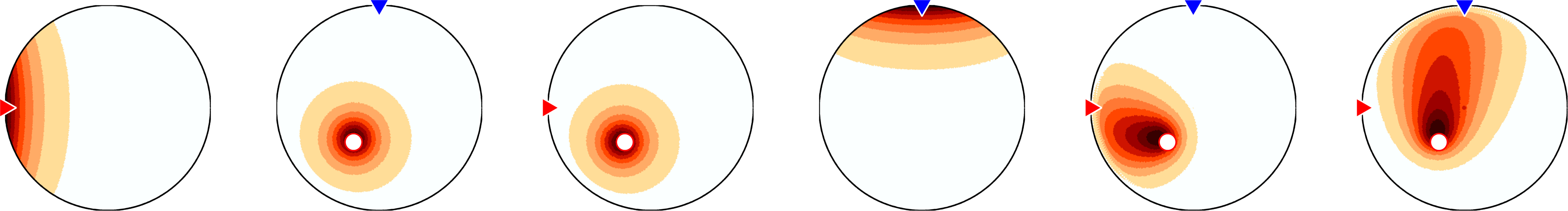}
\put(14.7, 6){\large $\times$}
\put(32, 6){\large $+$}
\put(49.5, 6){\large $\times$}
\put(66.5, 6){\large $=$}
\put(83.9, 6){\large $+$}
\end{overpic}
\caption{Pictorial description of eq.\ (\ref{eq:cdescmua}). From left to right:  $\bm{\phi}$, $\bm{\phi}_1^+$, $\bm{\phi}_1$, $\bm{\phi}^+$, $\bm{\phi} \times \bm{\phi}_1^+$, $\bm{\phi}_1 \times \bm{\phi}^+$. The final summation is represented in the CMDF depicted in column three of figure \ref{fig:cmdfsall}.
Red inwards arrows indicates real source positions, blue inwards arrows indicate adjoint source positions. White dots indicates ultrasound field focal point. The colour scale varies between plots.}
\label{fig:cmdfcomponents}
\end{figure*}
\begin{figure*}[!htb]
\centering
\begin{overpic}{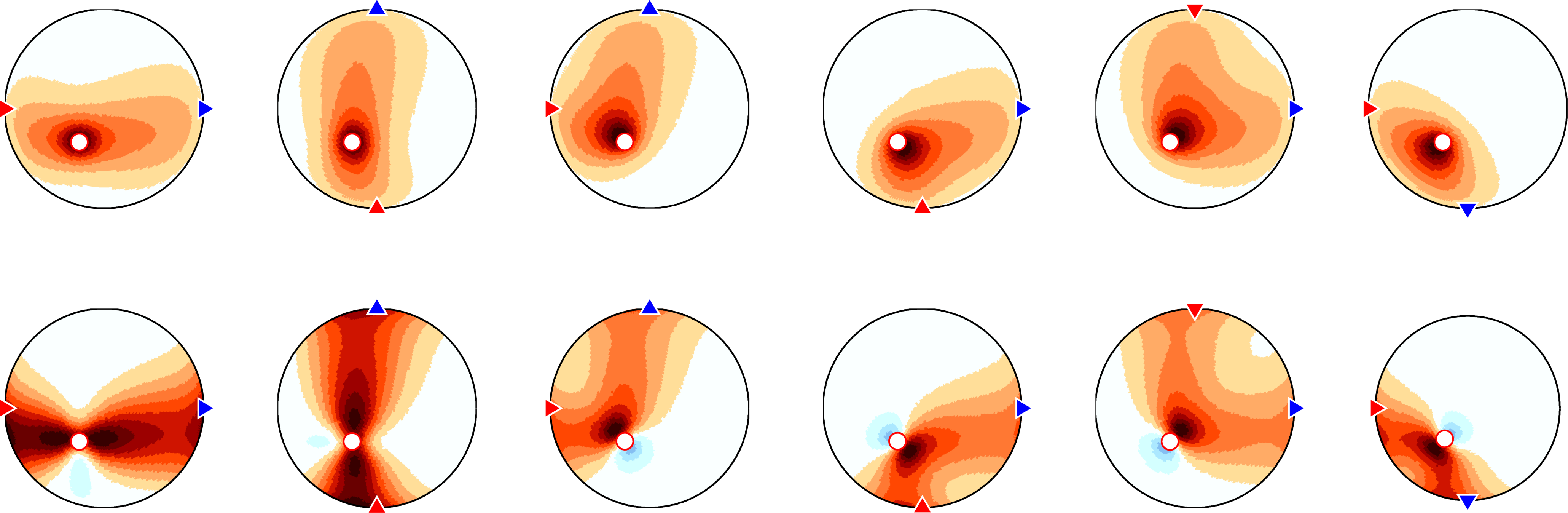}
\put(-4, 25){$\mua$}
\put(-4, 5.5){$\musp$}
\put(0,16){\line(50,0){100}}
\end{overpic}
\caption{A set of CMDFs in $\mua$ (top) and $\musp$ (bottom) for all optical source-detector pairs one (left) to six (right). Red arrows indicate source locations, blue arrows indicate detector locations, white dots indicate the focal point of the acoustic field. Red regions indicate that increases in perturbations cause reductions in the measured data, blue regions indicate the converse. The colour scale vary between plots.}
\label{fig:cmdfsall}
\end{figure*}
In figure \ref{fig:cmdfcomponents} we plot each of these terms for a specific acoustic field distribution and optical source-detector pair. The sensitivity function in $\mu_a$ consists of the summation of two products, which we also depict. The first product $\bm{\phi} \times \bm{\phi}_1^+$ (depicted in the fifth image of figure \ref{fig:cmdfcomponents}) represents the sensitivity of the  measurement due to attenuation of the input illumination which generates the first harmonic modulated fluence. We see a peak sensitivity in the region of the acoustic focus: this is partially due to our choice of a distributed optical source. Evidently if the optical source were point-like, the sensitivity to a perturbation in $\mua$ near this point could potentially be larger than that in the region of the distributed acoustic field, since a reduction in the strength of the optical source will cause a linear change in the total fluence. The second product $\bm{\phi}_1 \times \bm{\phi}^+$ (depicted in the sixth image of figure \ref{fig:cmdfcomponents}) represents the sensitivity of measurement due to attenuation of the modulated field prior to detection. As before, the exact form of this term is determined by our choice of detector profile. In summation, these two terms represent the sensitivity of a measurement of the modulated fluence due to a perturbation in the optical absorption profile. These figures demonstrate significant similarities to those derived for the non-linearised form of the forward model which we investigated in \cite{Powell:2012jb}. However the use of a point source and small aperture detector in the referenced work lead to increased sensitivity near the source and detector regions (we will return to this point in the discussion).

In figure \ref{fig:cmdfsall} we plot the CMDFs for $\mua$ and $\musp$ which arise for all six source detector pairs and a given ultrasound field distribution. The form of the sensitivity functions of the absorption coefficient follows our previous exposition; in each case a region of peak sensitivity is seen near the acoustic focus, extending outwards towards the given source and detector position for that measurement. The form of the sensitivity functions for $\musp$ retain the obvious dependence on the source and detector location, but have a more complicated structure in the region of the acoustic focus by virtue of their dependence upon the divergence of the four fields from which they are generated. In particular, we see a (spatially) fast reduction in sensitivity in the region where the forward and adjoint optical sensitivities are shadowed by the acoustic focus. We find areas of negative sensitivity in the shadowed region, where an increase in scattering will cause more of the input light to be modulated and detected.

\subsection{Two-dimensional reconstruction}
\label{res:2drecon}

\begin{figure}[b!]
\centering
\begin{overpic}{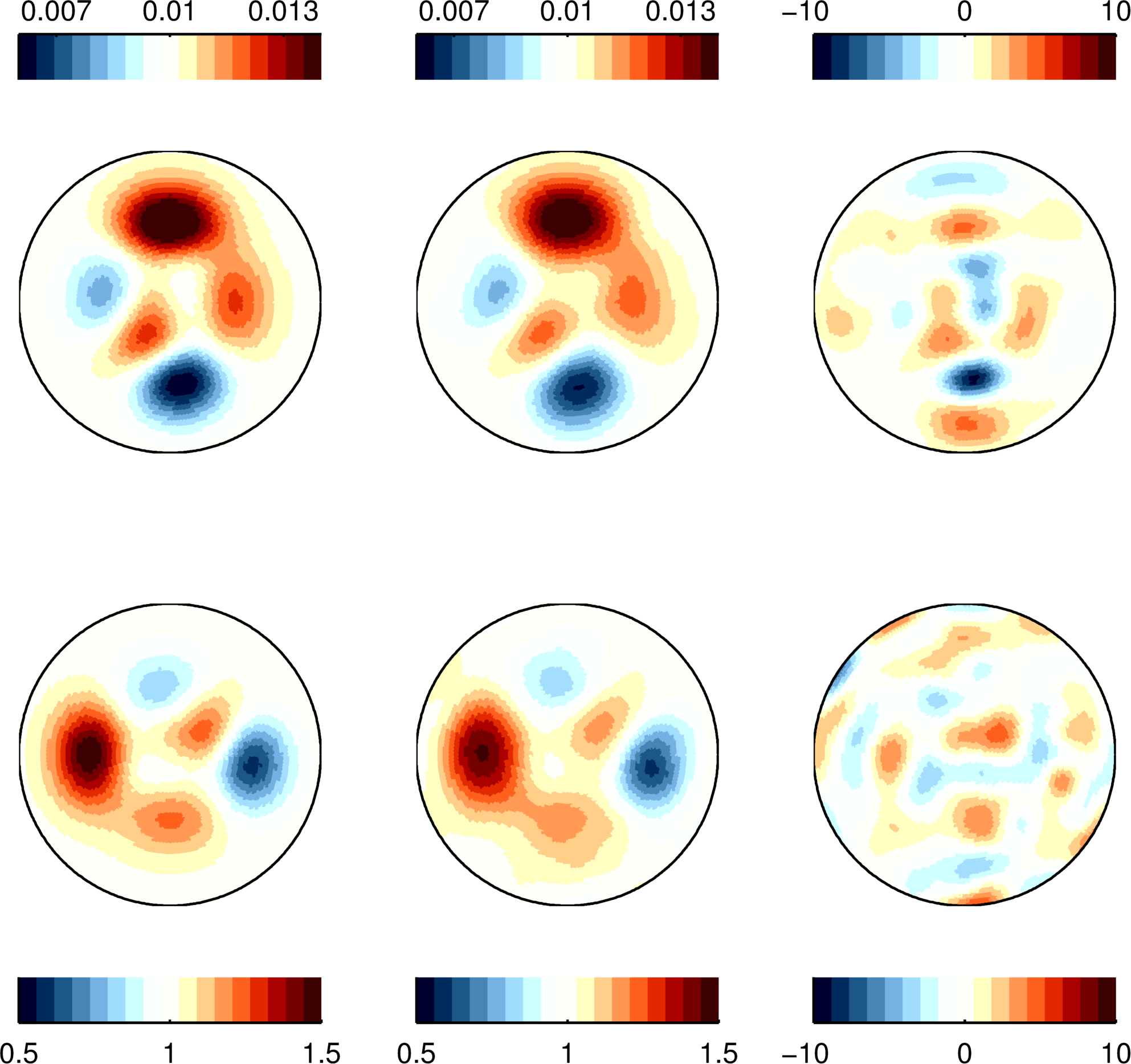}
\put(-3.5, 65){\small{$\mua$}}
\put(-3.5, 25){\small{$\musp$}}
\put(5,46){\small{Target [mm$^{-1}$]}}
\put(34,46){\small{Reconstruction [mm$^{-1}$]}}
\put(79,46){\small{Error [\%]}}
\end{overpic}
\caption{Target (left), reconstruction (middle) and percentage error (right) images of $\mua$ (top) and $\musp$ (bottom) for two-dimensional reconstruction.}
\label{fig:recon2d}
\end{figure}
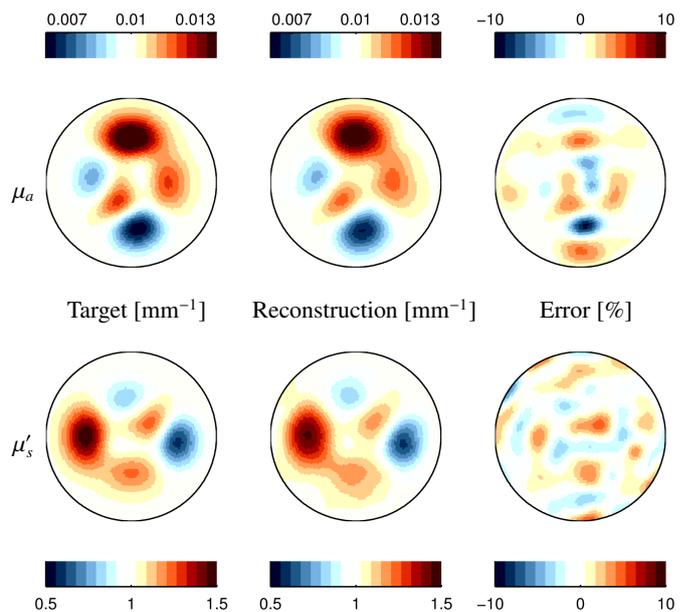

To demonstrate our algorithm in the reconstruction of an image we introduce perturbations in the absorption and scattering coefficients of the two-dimensional domain utilised in section \ref{res:cmdfs}. Simulated measurements were performed for all source-detector pairs and $1\%$ proportional Gaussian noise added to the data. The error covariance matrix $\ige$ was set to diag$(1/y)$ in accordance with our noise model. The regularisation hyper-parameter was selected by inspection.  The optimisation was performed by the non-linear conjugate gradient method, and was terminated when the change in the objective function fell below $\Delta \defunc \leq 0.01$, which resulted in convergence after 26 iterations. Figure \ref{fig:recon2d} depicts the target optical parameters, the results of our reconstruction, and the percentage error for each coefficient.
The reconstructed images can be seen to be in excellent agreement with their associated targets. The error in $\mua$ appears to be correlated with the image insofar as there is a slight under-reporting of the peak-positive and negative perturbations. The error does not exceed $\pm 5\%$ at any point in the image. A similar result is seen in the error in the reconstruction of $\musp$.

\subsection{Three-dimensional reconstruction}
\label{sec:recon3d}

\begin{figure}[b!]
\centering
\includegraphics{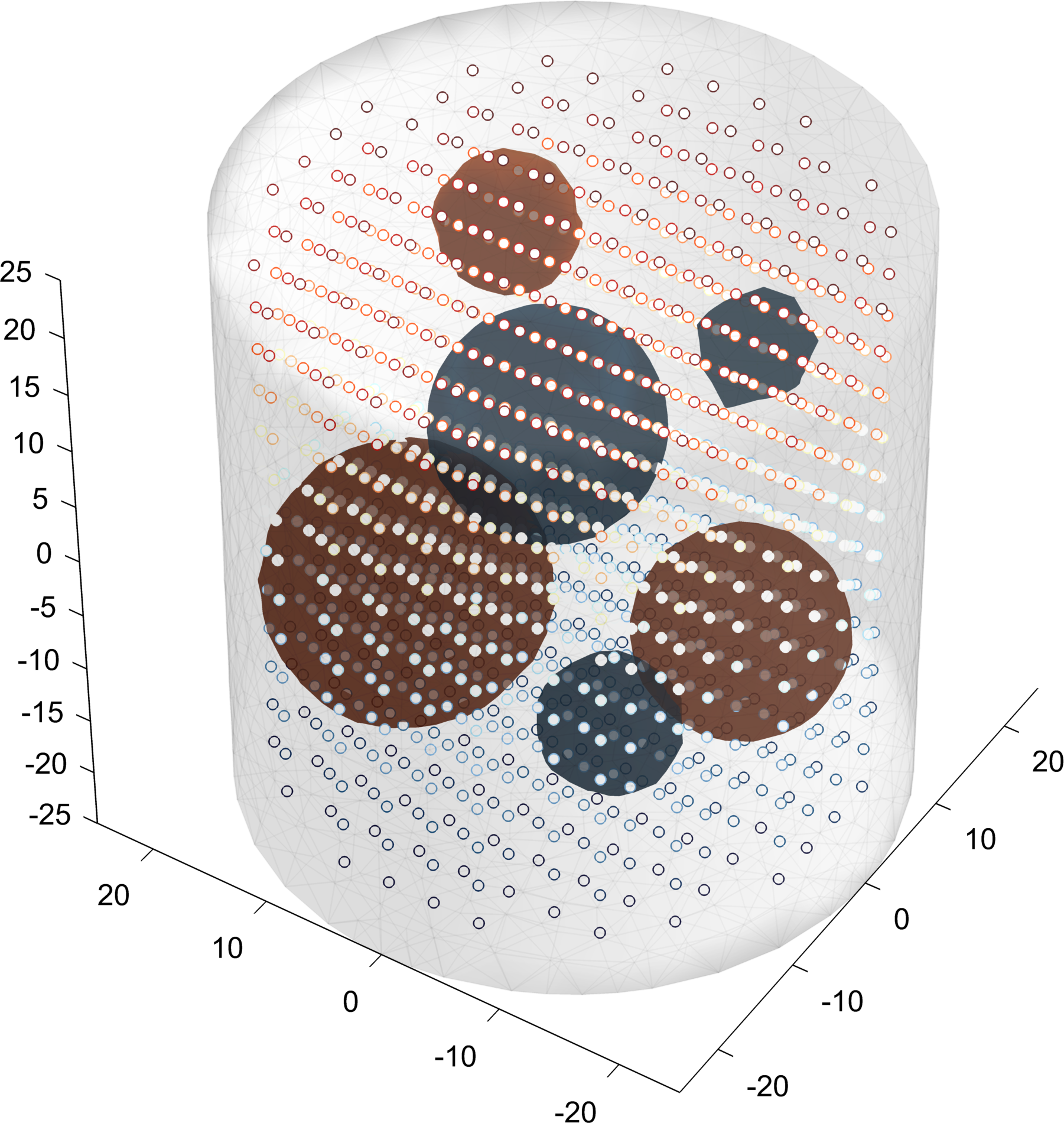}
\caption{Geometry of the three-dimensional mesh for the reconstructions of section \ref{sec:recon3d}. Blue volumes indicate scattering perturbations, red volumes indicate absorption perturbations, coloured dots indicate the location of the acoustic focii, where the colour indicates the location of the focus in the $z$-axis for easier visualisation.}
\label{fig:mesh3dfields}
\end{figure} 
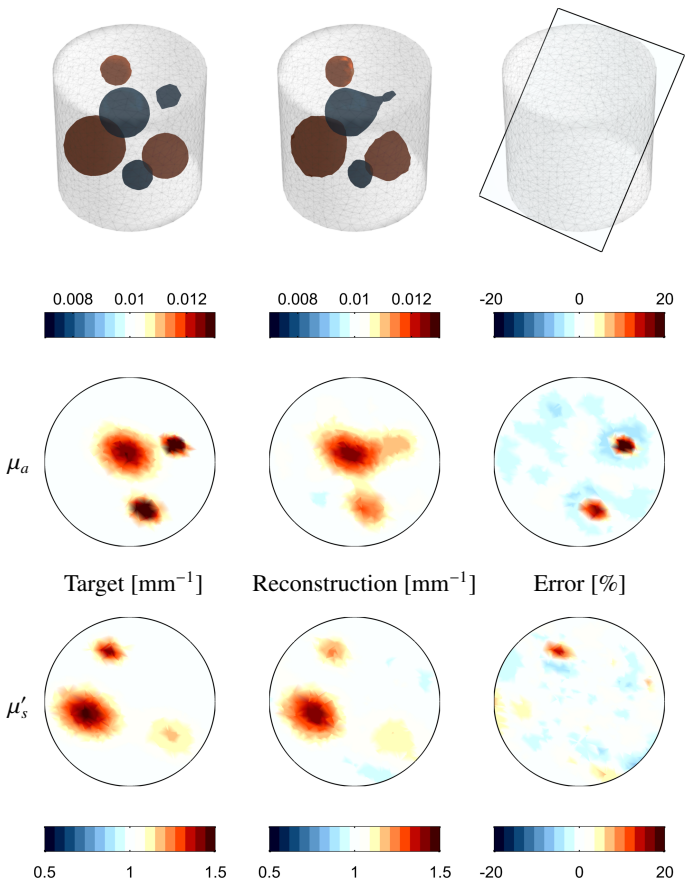
\begin{figure}[b!]
\centering
\begin{overpic}{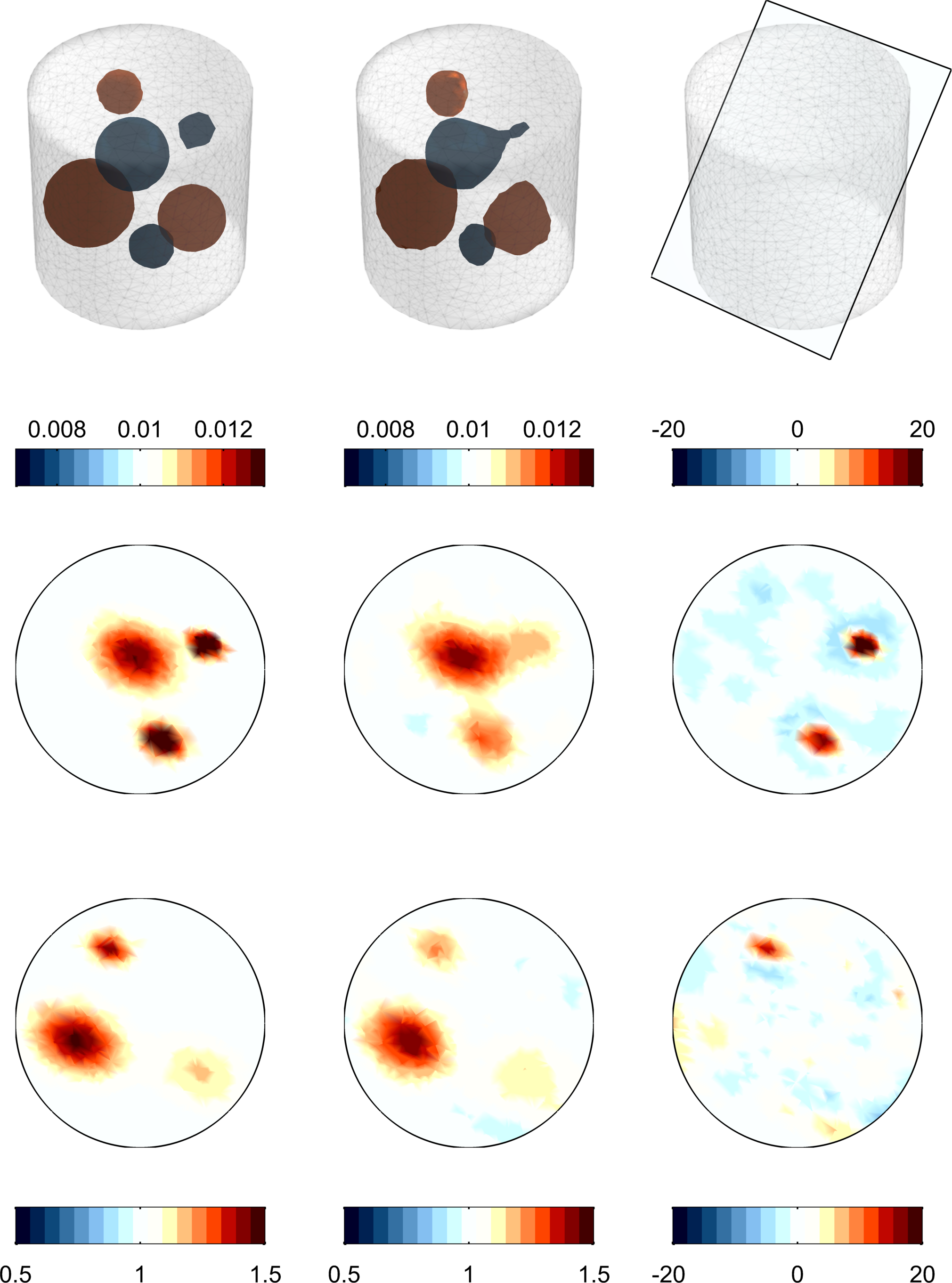}
\put(-3, 47){\small{$\mua$}}
\put(-3, 19){\small{$\musp$}}
\put(3.42,33){\small{Target [mm$^{-1}$]}}
\put(24.9,33){\small{Reconstruction [mm$^{-1}$]}}
\put(57,33){\small{Error [\%]}}
\end{overpic}
\caption{Isosurfaces of the target (top left), and reconstructed (top middle) parameter distributions, red indicates perturbations in $\mua$, blue indicates $\musp$. The top right figure indicates the plane through which the projections of $\mua$ and $\musp$ in the lower rows is taken. Middle and bottom rows illustrate the target (left) and reconstructed (middle) parameters, and the percentage error (right), for $\mua$ (middle row), and $\musp$ (bottom row).}
\label{fig:recon3d}
\end{figure}

To demonstrate a three-dimensional reconstruction we consider a cylindrical domain of diameter and height $50\si{\mm}$. The absorption and scattering coefficients consist of an homogeneous background $\mua = 0.01\si{\per\mm}$, $\musp = 1\si{\per\mm}$, with numerous perturbations in both parameters with a maximum magnitude of $50\%$ of the background value. A set of 1456 focused acoustic fields with three-dimensional Gaussian profiles of FWHM = $3.5\si{\mm}$ probe the domain. The focal points are arranged over a rectangular grid, truncated at a radius of $22\si{\mm}$. The peak magnitude of $\eta(\vr) = 0.25$. The geometry of the problem, the location of the focal points of the ultrasound fields, and isosurfaces of the optical parameter perturbations are depicted in figure \ref{fig:mesh3dfields}. Four sources and detectors were arranged around the periphery of the domain. The first three sources and detectors were located as per the two-dimensional problem, at $z=0$. The final source and detector were placed on the top and bottom surfaces of the domain, respectively. Each source had a profile corresponding to a  cosine window of diameter $20\si{mm}$. Simulated measurements were performed for all source detector pairs and $1\%$ proportional Gaussian noise was added to the data. The error covariance matrix and regularisation hyper-parameter were determined as per the two-dimensional case. The optimisation was performed using the preconditioned non-linear conjugate gradient method, and was terminated when the change in the objective function fell below $\Delta \defunc \leq 0.01$, which resulted in convergence after 61 iterations. Figure \ref{fig:recon3d} compares the target optical parameters, the results of the reconstruction, and the percentage error for each coefficient for a slice through the domain.

The reconstructed images successfully capture the location and magnitude of the larger perturbations in the optical coefficients of the domain with excellent accuracy. The smaller perturbations are somewhat under-reported, especially in the case of the absorption coefficient. 

This can be directly attributed to the use of a coarser grid of ultrasound focal points, and the increased FWHM of the fields, which in this case is comparable to the size of the smaller inclusions. As such, this system is significantly more under-determined than the two-dimensional case, as the number of degrees of freedom has grown significantly compared to the number of measurements (ultrasound field locations).

The result of this is that the weighting of the prior, which in this case enforces smoothness, is relatively larger. This leads to the small absorption perturbation `merging' into the larger feature, which can be seen both in the isosurface and slice plots. An edge preserving prior such as total variation may better retain delineation between regions of differing optical properties.

\section{Discussion \& Conclusions}
\label{sec:conclusion}
Before closing, we consider some points raised by this work, and make suggestions for future investigations.

\subsection{The coherent optical source, and uniqueness}

We noted earlier that the use of a diffuse source term of similar spatial extent to the detector profile resulted in UOT sensitivity functions with a maximum sensitivity at the point of the acoustic field, and that this differed from previous work \cite{Powell:2013dx} in which a point source dominated the sensitivity of the system for this measurement type. In another recent publication which employed point sources \cite{Powell:2014eu} we noted that interchanging of the source and detector locations lead to an improvement in the spectrum (the number of significant singular values) of the approximated Hessian which arises in a quasi-Newton optimisation: this has significant implications regarding the uniqueness of the reconstruction. This previously unexplained phenomenon seems at odds with the symmetry which results from the physical reciprocity of the system. 

In this work, with identical source and measurement apertures, no advantages were gained from taking measurements with swapped source-detector locations. Thus, the extra information found by transposing the source and detectors in \cite{Powell:2014eu} was in fact achieved by virtue of their differing profiles.
Whilst this might initially suggest an advantage in using a point source and diffuse detector, we must consider that the extreme sensitivity near the source position may be deleterious: not only will the increased dynamic range of the sensitivity function lead to a deterioration of the condition number of matrices to be inverted, but any experimental system will become highly sensitive to miss-location of the applied optical sources. Moreover, a greater amount of total optical power can be delivered to a tissue experimentally if it is illuminated by a diffuse large aperture source.

In this work we gained sufficiently independent sets of internal data to permit the simultaneous recovery of $\mua$ and $\musp$ by the use of multiple optical sources and detectors, but this may be undesirable in practice. It is not at present evident  the way in which independence in the interior data can best be achieved (the choice of multiple optical source and detector locations signifies the implicit assumption that the independence is fundamentally related to the spatial gradient of the internal fields). This point is worthy of further consideration, since if independence could be achieve by another means, for example, an appropriate set of structured illumination patterns, this would have significant experimental advantages. 

\subsection{Measurement types, and noise}

In this work we presented the use of the first-harmonic fluence as our measurement type. This data-type arises naturally in a number of UOT detection mechanisms. In applying these reconstruction techniques to experimental data the forward model will need to be normalised in some way to match the experimental observations. Like in the case of DOT, these `coupling-coefficients' must remain constant through the experiment, lest significant error be introduced to the reconstruction.

UOT offers a more robust measurement type, that of the modulation depth, which in our linear formulation would be found as the ratio of the first-harmonic fluence to the total fluence. This measurement-type has the attractive advantage of being self-normalised, such that changes in the coupling coefficients, providing they are consistent across the power spectrum, will not affect the measurement. We employed a modulation depth measurement type in a lag-domain model presented in a previous work \cite{Powell:2013dx}, and found the sensitivity functions to demonstrate improved spatial localisation, and complete suppression of the optical source and detector locations. Similar results were found experimentally by Gunadi and Leung \cite{Gunadi:2011ch}. This was manifested in improved reconstructions which were insensitive to perturbation near the source location, even when a point-source was employed. This is of importance in various clinical applications where superficial changes in blood volume during, for example, functional response, or therapy, may otherwise come to dominate an image. For these reasons, it would be of significant value to extend the analysis of this work to the modulation depth measurement type.

For generality, we chose to apply proportional Gaussian noise to our simulated measurements. This choice is equivalent to shot noise in the limit of large signal levels and equal intensity at each detector \cite{Arridge:2009cy}: a fair model for interferometric detection systems. Alternative noise models will be required if these techniques are to be applied to true photon-counting systems, appropriate examples can be found in the literature of DCS \cite{Zhou:2006uw}.

\subsection{Towards application}

UOT is a young imaging modality, and there are currently many different methods by which coherent light is measured: each one has different constraints on the geometries to which it can be applied, and more significantly, implies different noise characteristics in the data. No single method has yet proven itself to be the obvious candidate for the future development of clinical technologies. The techniques we have developed in this work are of sufficient generality to be applied to a wide range of detection methods, but the performance that can be expected, viz. spatial resolution, noise immunity, and the ability to recover both absorption and scattering, will depend heavily on the specific experimental configuration.

\subsection{Summary}

In this work we have provided an overview of forward modelling techniques in UOT. We have demonstrated that a computationally efficient frequency-domain linearised diffusion model can be found by the formal linearisation of a diffusion approximation to a correlation transport equation for UOT. We have derived and elucidated the form of the correlation measurement density functions which arise for the first-harmonic measurement type. These sensitivity functions were employed in the derivation of the gradient of the an objective function. Employing this gradient in a non-linear optimisation technique permitted the simultaneous reconstruction of images of the optical absorption and scattering coefficients in both two- and three-dimensions.

\section*{Acknowledgements}
The authors would like to thank Ben Cox, Jem Hebden, and Emma Malone for their advice and support. This work was partially funded by EPSRC grant EP/G005036/1.

\bibliographystyle{elsarticle-num}
\bibliography{./bib/gradquot}

\begin{thebibliography}{10}
\expandafter\ifx\csname url\endcsname\relax
  \def\url#1{\texttt{#1}}\fi
\expandafter\ifx\csname urlprefix\endcsname\relax\def\urlprefix{URL }\fi
\expandafter\ifx\csname href\endcsname\relax
  \def\href#1#2{#2} \def\path#1{#1}\fi

\bibitem{Arridge:2009cy}
S.~R. Arridge, J.~C. Schotland, {Optical tomography: forward and inverse
  problems} 25~(12) (2009) 123010.

\bibitem{Arridge:1999kd}
S.~R. Arridge, {Optical tomography in medical imaging}, Inverse Problems 15~(2)
  (1999) R41--R93.

\bibitem{Gibson:2005hr}
S.~R. Arridge, A.~P. Gibson, J.~C. Hebden, {Recent advances in diffuse optical
  imaging}, Physics in Medicine and Biology 50~(4) (2005) R1--R43.

\bibitem{Gross:2003ei}
M.~Gross, M.~Al-Koussa, {Shot-noise detection of ultrasound-tagged photons in
  ultrasound-modulated optical imaging}, Optics Letters 28~(24) (2003)
  2482--2484.

\bibitem{Leveque:1999tm}
S.~L{\'e}v{\^e}que, A.~C. Boccara, {Ultrasonic tagging of photon paths in
  scattering media: parallel speckle modulation processing}, Optics Letters.

\bibitem{Li:2002wq}
J.~Li, L.~V. Wang, G.~Ku, {Ultrasound-modulated optical tomography of
  biological tissue by use of contrast of laser speckles.}, Applied Optics
  41~(28) (2002) 6030--6035.

\bibitem{Ramaz:2004vs}
A.~C. Boccara, F.~Ramaz, B.~Forget, M.~Atlan, M.~Gross, P.~Delaye, G.~Roosen,
  {Photorefractive detection of tagged photons in ultrasound modulated optical
  tomography of thick biological tissues.}, Optics Express 12~(22) (2004)
  5469--5474.

\bibitem{Murray:2004tb}
F.~J. Blonigen, T.~W. Murray, L.~Sui, G.~Maguluri, R.~A. Roy, A.~Nieva, C.~A.
  DiMarzio, {Detection of ultrasound-modulated photons in diffuse media using
  the photorefractive effect.}, Optics Letters 29~(21) (2004) 2509--2511.

\bibitem{Bossy:2005tr}
E.~Bossy, L.~Sui, T.~W. Murray, R.~A. Roy, {Fusion of conventional ultrasound
  imaging and acousto-optic sensing by use of a standard pulsed-ultrasound
  scanner.}, Optics Letters 30~(7) (2005) 744--746.

\bibitem{Li:2008hy}
L.~V. Wang, Y.~Li, H.~Zhang, C.~Kim, K.~H. Wagner, P.~Hemmer, {Pulsed
  ultrasound-modulated optical tomography using spectral-hole burning as a
  narrowband spectral filter.}, Applied Physics Letters 93~(1) (2008) 11111.

\bibitem{Suzuki:2013ft}
L.~V. Wang, Y.~Suzuki, P.~Lai, X.~Xu, {High-sensitivity ultrasound-modulated
  optical tomography with a photorefractive polymer}, Optics Letters 38~(6)
  (2013) 899--901.

\bibitem{Li:2002jg}
L.~V. Wang, H.~Li, {Autocorrelation of Scattered Laser Light for
  Ultrasound-Modulated Optical Tomography in Dense Turbid Media}, Applied
  Optics 41~(22) (2002) 4739--4742.

\bibitem{Lev:2002ba}
B.~G. Sfez, A.~Lev, {Direct, noninvasive detection of photon density in turbid
  media}, Optics Letters 27~(7) (2002) 473.

\bibitem{Lev:2003dt}
B.~G. Sfez, A.~Lev, {Pulsed ultrasound-modulated light tomography}, Optics
  Letters 28~(17) (2003) 1549--1551.

\bibitem{Arridge:1995fy}
S.~R. Arridge, {Photon-measurement density functions. Part I: Analytical
  forms}, Applied Optics 34~(31) (1995) 7395--7409.

\bibitem{Allmaras:kr}
M.~Allmaras, W.~Bangerth, {Reconstructions in ultrasound modulated optical
  tomography}, Journal of Inverse and Ill-posed Problems 19 (2011) 801--823.

\bibitem{Powell:2013dx}
S.~Powell, T.~S. Leung, {Linear reconstruction of absorption perturbations in
  coherent ultrasound-modulated optical tomography}, Journal of Biomedical
  Optics 18~(12) (2013) 126020--126020.

\bibitem{Bratchenia:2011ch}
A.~Bratchenia, R.~Molenaar, T.~G. van Leeuwen, R.~P.~H. Kooyman,
  {Acousto-optic-assisted diffuse optical tomography}, Optics Letters 36~(9)
  (2011) 1539.

\bibitem{Bal:2010gv}
G.~Bal, J.~C. Schotland, {Inverse Scattering and Acousto-Optic Imaging},
  Physical Review Letters 104~(4) (2010) 043902.

\bibitem{Bal:2014cg}
G.~Bal, S.~Moskow, {Local inversions in ultrasound-modulated optical
  tomography}, Inverse Problems 30~(2) (2014) 025005.

\bibitem{Powell:2014eu}
S.~Powell, T.~S. Leung, {Quantitative reconstruction of absorption and
  scattering coefficients in coherent ultrasound-modulated optical tomography},
  SPIE BiOS 8943 (2014) 89434X--89434X--11.

\bibitem{Beard:2011bm}
P.~C. Beard, {Biomedical photoacoustic imaging}, Interface Focus 1~(4) (2011)
  602--631.

\bibitem{Kothapalli:2007hx}
S.-R. Kothapalli, S.~Sakad{\v z}i{\'c}, C.~Kim, L.~V. Wang, {Imaging optically
  scattering objects with ultrasound-modulated optical tomography}, Optics
  Letters 32~(16) (2007) 2351--2353.

\bibitem{Wang:2008ci}
L.~V. Wang, {Prospects of photoacoustic tomography}, Medical Physics 35~(12)
  (2008) 5758--5767.

\bibitem{BenCox:2012jr}
B.~T. Cox, J.~G. Laufer, S.~R. Arridge, P.~C. Beard, {Quantitative
  spectroscopic photoacoustic imaging: a review}, Journal of Biomedical Optics
  17~(6) (2012) 0612021--06120222.

\bibitem{Gao:2012cu}
H.~Gao, S.~Osher, H.~Zhao, {Quantitative Photoacoustic Tomography},
  Mathematical Modeling in Biomedical Imaging II (2012) 131--158.

\bibitem{Saratoon:2013bm}
S.~R. Arridge, T.~Saratoon, T.~Tarvainen, B.~T. Cox, {A gradient-based method
  for quantitative photoacoustic tomography using the radiative transfer
  equation}, Inverse Problems 29~(7) (2013) 075006.

\bibitem{Hochuli:2015et}
R.~Hochuli, S.~Powell, S.~R. Arridge, B.~T. Cox, { Forward and adjoint radiance
  Monte Carlo models for quantitative photoacoustic imaging}, SPIE BiOS 9323
  (2015) 93231P--93231P--10.

\bibitem{Wang:2001jd}
L.~V. Wang, {Mechanisms of Ultrasonic Modulation of Multiply Scattered Coherent
  Light: An Analytic Model}, Physical Review Letters 87~(4) (2001) 043903.

\bibitem{Leutz:1995hf}
W.~Leutz, G.~Maret, {Ultrasonic modulation of multiply scattered light},
  Physica B: Condensed Matter 204~(1-4) (1995) 14--19.

\bibitem{Kempe:1997ig}
M.~Kempe, M.~Larionov, D.~Zaslavsky, A.~Z. Genack, {Acousto-optic tomography
  with multiply scattered light}, Journal of the Optical Society of America A
  14~(5) (1997) 1151--1158.

\bibitem{Elson:tm}
D.~S. Elson, R.~Li, C.~Dunsby, R.~Eckersley, {Ultrasound-mediated optical
  tomography: a review of current methods}, Interface Focus.

\bibitem{Maret:1987vl}
G.~Maret, P.~E. Wolf, {Multiple light scattering from disordered media. The
  effect of brownian motion of scatterers}, Zeitschrift f{\"u}r Physik B
  Condensed Matter 65~(4) (1987) 409--413.

\bibitem{Pine:1988vp}
D.~J. Pine, D.~A. Weitz, P.~M. Chaikin, E.~Herbolzheimer, {Diffusing wave
  spectroscopy}, Physical Review Letters 60~(12) (1988) 1134--1137.

\bibitem{Sakadzic:2002bj}
S.~Sakad{\v z}i{\'c}, L.~V. Wang, {Ultrasonic modulation of multiply scattered
  coherent light: an analytical model for anisotropically scattering media},
  Physical Review E.

\bibitem{Sakadzic:2005go}
L.~V. Wang, S.~Sakad{\v z}i{\'c}, {Modulation of multiply scattered coherent
  light by ultrasonic pulses: An analytical model}, Physical Review E 72~(3)
  (2005) 036620.

\bibitem{Ackerson:1992fx}
B.~J. Ackerson, R.~L. Dougherty, N.~M. Reguigui, U.~Nobbmann, {Correlation
  transfer: Application of radiative transfer solution methods to photon
  correlation problems}, Journal of thermophysics and heat transfer 6~(4)
  (1992) 577--588.

\bibitem{Dougherty:1994vn}
R.~L. Dougherty, B.~J. Ackerson, N.~M. Reguigui, F.~Dorri-Nowkoorani,
  U.~Nobbmann, {Correlation transfer: Development and application}, Journal of
  Quantitative Spectroscopy and Radiative Transfer 52~(6) (1994) 713--727.

\bibitem{Ishimaru:1975uv}
A.~Ishimaru, S.~T. Hong, {Multiple scattering effects on coherent bandwidth and
  pulse distortion of a wave propagating in a random distribution of
  particles.}, Radio Science 10~(6) (1975) 637--644.

\bibitem{Stephen:1988in}
M.~Stephen, {Temporal fluctuations in wave propagation in random media},
  Physical Review B 37~(1) (1988) 1--5.

\bibitem{MacKintosh:1989jz}
F.~MacKintosh, S.~John, {Diffusing-wave spectroscopy and multiple scattering of
  light in correlated random media}, Physical Review B 40~(4) (1989)
  2383--2406.

\bibitem{Sakadzic:2007is}
S.~Sakad{\v z}i{\'c}, L.~V. Wang, {Correlation transfer equation for multiply
  scattered light modulated by an ultrasonic pulse}, Journal of the Optical
  Society of America A 24~(9) (2007) 2797.

\bibitem{Sakadzic:2006ei}
S.~Sakad{\v z}i{\'c}, L.~V. Wang, {Correlation transfer equation for
  ultrasound-modulated multiply scattered light}, Physical Review E (2006)
  036618--1 -- 036618--10.

\bibitem{Leung:2010jr}
T.~S. Leung, S.~Powell, {Fast Monte Carlo simulations of ultrasound-modulated
  light using a graphics processing unit}, Journal of Biomedical Optics 15~(5)
  (2010) 055007--055007--7.

\bibitem{Powell:2012jb}
S.~Powell, T.~S. Leung, {Highly parallel Monte-Carlo simulations of the
  acousto-optic effect in heterogeneous turbid media}, Journal of Biomedical
  Optics 17~(4) (2012) 045002--04500211.

\bibitem{Boas:1997kf}
D.~A. Boas, A.~G. Yodh, {Spatially varying dynamical properties of turbid media
  probed with diffusing temporal light correlation}, Journal of the Optical
  Society of America A 14~(1) (1997) 192.

\bibitem{Sakadzic:2006tx}
L.~V. Wang, S.~Sakad{\v z}i{\'c}, {Correlation transfer and diffusion of
  ultrasound-modulated multiply scattered light}, Physical Review Letters
  96~(16) (2006) 163902--1 --163902--4.

\bibitem{Schweiger:1995to}
M.~Schweiger, S.~R. Arridge, M.~Hiraoka, D.~T. Delpy, {The finite element
  method for the propagation of light in scattering media: boundary and source
  conditions}, Medical Physics 22~(11) (1995) 1779--1792.

\bibitem{Arridge:1995ho}
S.~R. Arridge, M.~Schweiger, {Photon-measurement density functions. Part 2:
  Finite-element-method calculations}, Applied Optics 34~(34) (1995)
  8026--8037.

\bibitem{Arridge:1993gv}
S.~R. Arridge, {A finite element approach for modeling photon transport in
  tissue}, Medical Physics 20~(2) (1993) 299--309.

\bibitem{Heino:2003bd}
J.~Heino, S.~R. Arridge, J.~Sikora, E.~Somersalo, {Anisotropic effects in
  highly scattering media}, Physical Review E 68~(3) (2003) 031908.

\bibitem{Schweiger:2014ee}
M.~Schweiger, S.~R. Arridge, {The Toast++ software suite for forward and
  inverse modeling in optical tomography}, Journal of Biomedical Optics.

\bibitem{Arridge:1998wp}
S.~R. Arridge, M.~Schweiger, {A gradient-based optimisation scheme for optical
  tomography}, Optics Express 2~(6) (1998) 213--226.

\bibitem{Cox:2007he}
B.~T. Cox, P.~C. Beard, S.~R. Arridge, {Gradient-based quantitative
  photoacoustic image reconstruction for molecular imaging}, Biomedical Optics
  (BiOS) 2007 6437 (2007) 64371T--64371T--10.

\bibitem{Schweiger:2005fa}
M.~Schweiger, S.~R. Arridge, I.~Nissil{\"a}, {Gauss{\textendash}Newton method
  for image reconstruction in diffuse optical tomography}, Physics in Medicine
  and Biology 50~(10) (2005) 2365--2386.

\bibitem{Nocedal:428607}
J.~Nocedal, S.~J. Wright, {Numerical optimization}, Springer series in
  operations research, Springer, New York, NY, 1999.

\bibitem{Gunadi:2011ch}
S.~Gunadi, T.~S. Leung, {Spatial sensitivity of acousto-optic and optical
  near-infrared spectroscopy sensing measurements}, Journal of Biomedical
  Optics 16~(12) (2011) 127005--12700510.

\bibitem{Zhou:2006uw}
C.~Zhou, G.~Yu, D.~Furuya, J.~Greenberg, A.~Yodh, T.~Durduran, {Diffuse optical
  correlation tomography of cerebral blood flow during cortical spreading
  depression in rat brain.}, Optics Express 14~(3) (2006) 1125--1144.

\end{thebibliography}

\end{document}